\documentclass[9pt,twocolumn,twoside]{osajnl}
\newcommand{\la}{\langle}
\newcommand{\ra}{\rangle}
\newcommand{\shiki}[1]{Eq. (\ref{#1})}
\newcommand{\zu}[1]{Fig. \ref{#1}}

\journal{josab} % Choose journal (ao, aop, josaa, josab, ol, pr)

\setboolean{shortarticle}{false}
\title{Angular response of a triangular optical cavity analyzed by a linear approximation method}

\author[1,2,*]{Satoshi Tanioka}
\author[3]{Gui-guo Ge}
\author[4]{Keiko Kokeyama}
\author[5]{Masayuki Nakano}
\author[6]{Junegyu Park}
\author[7]{Kiwamu Izumi}

\affil[1]{The Graduate University for Advanced Studies (SOKENDAI), Mitaka, Tokyo 181-8588, Japan}
\affil[2]{National Astronomical Observatry oj Japan (NAOJ), Mitaka, Tokyo 181-8588, Japan}
\affil[3]{State Key Laboratory of Magnetic Resonance and Atomic and Molecular Physics,
Wuhan Institute of Physics and Mathematics, Chinese Academy of Sciences, West
No. 30, Xiaohongshan, Wuhan, 430071, China}
\affil[4]{Institute for Cosmic Ray Research (ICRR), The University of Tokyo, Higashi-Mozumi 238, Kamioka-cho, Hida-shi, Gifu 506-1205 Japan}
\affil[5]{Department of Physics, University of Toyama, Toyama City, Toyama 930-8555,
Japan}
\affil[6]{Department of Physics, Sogang University, One Sinsu-Dong, Mapo-Gu, Seoul, 121-742, Korea}
\affil[7]{Institute of Space and Astronautical Science, 3-1-1 Yoshinodai, Chuo ward, Sagamihara, Kanagawa 252-5210, Japan }

\affil[*]{Corresponding author: satoshi.tanioka@nao.ac.jp}

\begin{abstract}
A triangular optical cavity is often used as a mode cleaning cavity in precision laser interferometry such as gravitational
wave detectors.
An alignment sensing and control system for maintaining the alignment of a mode cleaning cavity with respect to the incoming laser beam is critical for detector’s performance.
Therefore, understanding the behavior of the angular response is vital to both design and commissioning test of the alignment control system.
We present a linear approximation approach which not only simplifies the computation but also provides a comprehensive picture of the angular response.
The observable degrees of freedom in a triangular cavity is discussed based on the linear approximation.
\end{abstract}

\setboolean{displaycopyright}{true}

\begin{document}

\maketitle

\section{Introduction}
Optical cavities are widely used in precision measurement, such as gravitational wave detection \cite{Aso2013} and optical lattice clock \cite{Ushijima2016}.
In gravitational wave detectors, one or more triangular cavities, called the input mode cleaners (IMCs), are employed. The IMCs provide three critical functionalities --- spatial mode cleaning, polarization selection and frequency stabilization \cite{Mueller2016, Nagano2002}.

\begin{figure}[t]
\centering
\includegraphics[width=120pt]{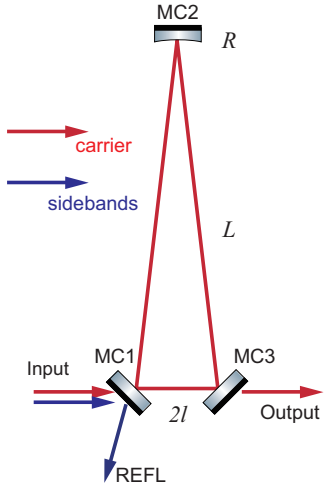}
\caption{Configuration of the IMC when the beam is injected to shorter side of the cavity. The IMC can be treated as a simple model since the carrier is on-resonant and the sidebands are off-resonant. The shorter side length $2l$ is much smaller than longer side length $L$. The apex mirror (MC2) has a radius of curvature, $R$. The beam waist located on halfway between the two flat mirrors, MCs 1 and 3.}
\label{IMC_top}
\end{figure}

The IMC is often an isosceles triangular cavity with the length of one side much shorter than the other two due to spatial constraint given by the vacuum envelope while maintaining tens of meters or more in the optical round trip length.
It is often the case that the triangular cavity is usually preferred because it does not require an additional optical isolator to extract the field in reflection such that its phase information is used to keep the cavity resonant.
The apex mirror is a curved mirror while the other two are flat mirrors as shown in Fig. \ref{IMC_top}.

Each mirror in the IMC is suspended to reduce seismic motions which, in turn, lead to angular drifts.
%input beam axis とcavity axisのズレ
Angular drifts directly introduce misalignment in the cavity axis. Such a misalignment can deteriorate detector sensitivity and stability by increasing a few noise couplings including those to length fluctuation and beam jitter.
Moreover, angular misalignment causes light in the fundamental mode to be coupled into higher-order spatial modes, reducing the amount of power injected to the main part of gravitational wave detector. This directly decreases the detector sensitivity.
Therefore, it is crucial to maintain the alignment of the IMC to achieve the detector operation with the best performance.
In order to reduce angular misalignment, an alignment sensing and control (ASC) system is adopted by using the wave front sensing (WFS) technique \cite{Morrison1994,Nergis1998}.

The WFS technique is sensitive to small tilts of the mirrors, and the WFS signals are obtained by demodulating signals detected by quadrant photodiodes (QPDs) \cite{Hefetz1997}.
Since the signals derived by the WFSs have a Gouy phase dependence, one has to precisely estimate its behavior and choose the QPD positions carefully in order to discriminate which mirror causes misalignments in either pitch or yaw.
Previous report shows the analytical calculation of the WFS signals \cite{Hefetz1997, Sigg2000}.
These approaches, however, contain complexities due to the fact that the non-diagonal matrix elements are fully incorporated in the calculation.
Therefore, the WFS signals of the IMC have been traditionally computed by simulation tools such as Finesse \cite{Freise2013Finesse} or Optickle \cite{Optickle}, and it was not easy to acquire intuitive interpretations.

In this article, we present a linear approximation method which enables us to relatively easily compute angular response of optical cavities.
We then show the angular responses of a triangular cavity such as the shift in beam spots and the WFS signals by employing the linear approximation method.
Finally, we will derive the sensing degrees of freedom (DoFs) using the singular value decomposition. Our study provides fundamental understanding with triangular optical cavities which is vital in the design and experimental phases.

\section{A Linear Approximation Method}
%In this section, the mathematical foundations for the linear approximation method are introduced.
%First, we define a number of conventions which are 
The $x$ and $y$ axes of the coordinate system are chosen to be transverse to the beam propagation (z axis) which is perfectly aligned. We additionally assume that the $y$ axis always points upwards regardless of the propagation direction. This automatically means that the $x$ axis needs to be mirrored every time the light is reflected by a mirror.
One can expand any paraxially approximated electromagnetic fields of light beams by a set of Hermit-Gaussian (HG) modes as \cite{Siegman}
\begin{align}
E(x,y,z) &= \sum_{lm}\la lm|E\ra U_{lm}(x,y,z), \\
\la lm|E\ra &\equiv \int_{-\infty}^{\infty}\int_{-\infty}^{\infty} U^*_{lm}(x,y,z)E(x,y,z)\mathrm{d}x\mathrm{d}y.
\end{align}
The coefficients $\la lm|E\ra$ can be represented as the elements of the vector in the modal space.
%misalignment excite HOMs
For small misalignment due to the mirror tilts, the only important modes are fundamental TEM00 mode and the second lowest-order transverse modes, TEM10 and TEM01 \cite{Hefetz1997}.
%Therefore, the field can be simplified as \color{red} This definition is not quite used in the discussion below. Necessary?\color{black}
%\begin{align}
%\vec{E} =
%\begin{pmatrix}
%\la00|E\ra \\ \la10|E\ra \\ \la01|E\ra
%\end{pmatrix}.
%\end{align}
Angular motions of a mirror in pitch correspond to rotations of the mirror about $x$ axis and excites a small amount of the TEM01 mode. Similarly, yaw is a rotation about the $r$ axis, exciting a TEM10 mode.
For simplicity, we assume that an angular misalignment exists only in pitch and leave TEM01 mode only into the consideration in the rest of this section. 

\begin{figure}[t]
\centering
\includegraphics[width=250pt]{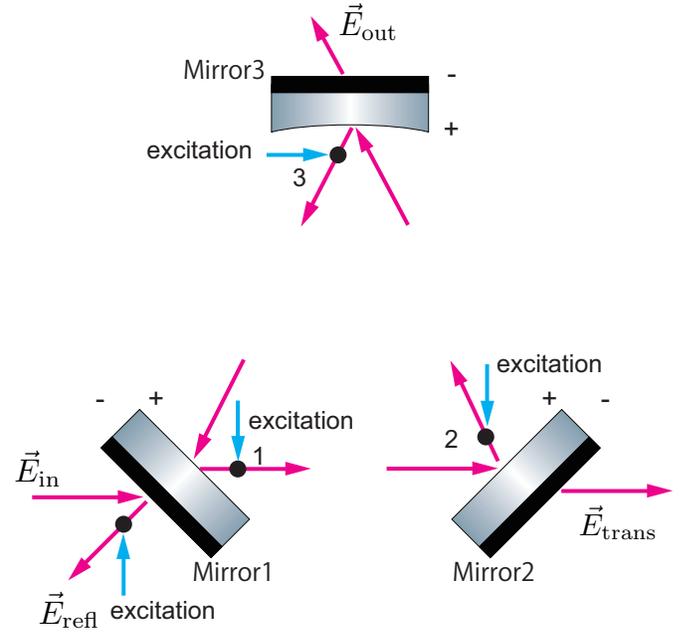}
\caption{The electric fields and nodes of a triangular cavity system. Red arrows and black dots represent electric fields and nodes, respectively. The misalignment of input mirror induces perturbation in the promptly reflection. The sign flip upon reflection of fields is labeled by "$+$" and "$-$" symbols.}
\label{node}
\end{figure}

We now consider a simple setup as shown in Fig. \ref{node} and shows how the linear approximation can be placed. Although a simple example is used, the argument here holds for any optical cavities as long as angular misalignment are sufficiently small. We define a collection of the field vectors containing the two sets of the fields i.e., one set from node 1 and the other from node 2.
\begin{align}
\vec{E}_{00}
\equiv
\begin{pmatrix}
\la 00 | E \ra_{1} \\ \la 00 | E \ra_{2} \\ \la 00 | E \ra_{3} \\ 
\end{pmatrix}.
\end{align}
These fields are related to each other and can be expressed in a matrix form~\cite{Freise2010},
\begin{align}
\begin{pmatrix}
0 & 0 & r_1\mathrm{e}^{-i\Phi_1} \\
r_2\mathrm{e}^{-i\Phi_{2}} & 0 & 0 \\
0 & r_3\mathrm{e}^{-i\Phi_3} & 0 \\
\end{pmatrix} 
&\begin{pmatrix}
\la 00 | E \ra_{1} \\ \la 00 | E \ra_{2} \\ \la 00 | E \ra_{3} \\ 
\end{pmatrix} \notag \\
+ 
\begin{pmatrix}
    t_1\la 00 | E \ra \\ 0 \\ 0 \\
    \end{pmatrix}
&= 
\begin{pmatrix}
\la 00 | E \ra_{1} \\ \la 00 | E \ra_{2} \\ \la 00 | E \ra_{3} \\ 
\end{pmatrix},
\label{field1}
\end{align}
where $\Phi_{i}$ $(i=1,2,3)$ is a one-way trip phase shift and $r_i$ are reflectivities of each mirror.
The incoming beam is assumed to be perfectly aligned to the cavity. In other words, only a TEM00 mode is pumped into the system.

To reduce clutter, we introduce a matrix and vectors as
\begin{align}
    \hat{M}_{00} &\equiv
    \begin{pmatrix}
    0 & 0 & r_1\mathrm{e}^{-i\Phi_1} \\
    r_2\mathrm{e}^{-i\Phi_{2}} & 0 & 0 \\
    0 & r_3\mathrm{e}^{-i\Phi_3} & 0 \\
    \end{pmatrix} , \\
    \vec{E}_{00,\mathrm{in}}
    &\equiv
    \begin{pmatrix}
    t_1\la 00 | E \ra \\ 0 \\ 0 \\
    \end{pmatrix}.
\end{align}
Then, one can rewrite Eq. (\ref{field1}) as
\begin{align}
    \hat{M}_{00}\vec{E}_{00} + \vec{E}_{00,\mathrm{in}} &= \vec{E}_{00}.
\end{align}
Similarly, the TEM01 mode fields satisfy the following relation,
\begin{align}
    \hat{M}_{01}\vec{E}_{01} &= \vec{E}_{01}, 
\end{align}
where
\begin{align}
    \vec{E}_{01}
    &\equiv
    \begin{pmatrix}
    \la 01 | E \ra_{1} \\ \la 01 | E \ra_{2} \\ \la 01 | E \ra_{3} \\
    \end{pmatrix}, \\
    \hat{M}_{01} &\equiv
    \begin{pmatrix}
    0 & 0 & r_1\mathrm{e}^{-i(\Phi_1+\eta_1)} \\
    r_2\mathrm{e}^{-i(\Phi_{2}+\eta_{2})} & 0 & 0 \\
    0 & r_3\mathrm{e}^{-i(\Phi_3+\eta_3)} & 0 \\
    \end{pmatrix}.
    %\vec{E}_{01,\mathrm{in}}
    %&\equiv
    %\begin{pmatrix}
    %t_1\la 01 | E \ra \\ 0 \\ 0 \\
    %\end{pmatrix}.
\end{align}
$\eta_{i}$ $(i=1,2,3)$ represents a Gouy phase shift.

Small angular misalignments are introduced in pitch which transfers a fraction of the TEM00 modes to TEM01 and vice versa \cite{Hefetz1997}.
We will express these perturbations using normalized rotation angles, $\Theta_i, (i=1,2)$.
These angles are normalized by divergence angle of the beam, $\lambda/\pi w(z)$, and expressed as
\begin{align}
\Theta_i &\equiv 2\frac{\pi w(z)}{\lambda}\theta_i,
\label{theta}
\end{align}
where the $\theta_i$ is a small rotation of each mirror, $\lambda$ is the wavelength of light and $w(z)$ is a beam size at a position of $z$ \cite{Hefetz1997}.
By using this normalized angle, the angular misalignment matrix is defined as
\begin{align}
    \hat{\Theta} &\equiv
    \begin{pmatrix}
    0 & 0 & -ir_1\mathrm{e}^{-i\Phi'_1}\Theta_1 \\
    -ir_2\mathrm{e}^{-i\Phi'_{2}}\Theta_3 & 0 & 0 \\
    0 & -ir_3\mathrm{e}^{-i\Phi'_3}\Theta_2 & 0 \\
    \end{pmatrix},
\end{align}
%wrong definition
%\begin{align}
    %\hat{\Theta} &\equiv
    %\begin{pmatrix}
    %0 & -ir_2\Theta_2 \\
    %-ir_1\Theta_1 & 0 \\
    %\end{pmatrix}.
%\end{align}
where $\Phi'_i$ represents the one-way trip phase shift for TEM00 or TEM01 modes.
Then one can express the fields with pitch angular misalignments as
\begin{align}
    \hat{M}_{00}\vec{E}_{00} + \vec{E}_{00,\mathrm{in}} + \hat{\Theta}\vec{E}_{01} &= \vec{E}_{00}, \label{E00} \\
    \hat{M}_{01}\vec{E}_{01} + \hat{\Theta}\vec{E}_{00} &= \vec{E}_{01}.
\end{align}
Since the incoming beam is pure TEM00 mode, TEM01 mode only exists when there is an angular misalignment.
The fields of TEM01 mode can be written as
\begin{align}
    \vec{E}_{01} &= (I-\hat{M}_{01})^{-1}\hat{\Theta}\vec{E}_{00} \notag \\
    &= (I-\hat{M}_{01})^{-1}\hat{\Theta}(I-\hat{M}_{00})^{-1}(\vec{E}_{00,\mathrm{in}}+\hat{\Theta}\vec{E}_{01}),
    \label{E01}
\end{align}
where $I$ is the identity matrix
Here we substituted $\vec{E}_{00}$ given by Eq. (\ref{E00}).
Since the angular misalignment angle is so small that one can ignore higher order term $O(\hat{\Theta}^2)$. In fact, this is the essence of our linear approximation.
With this linear approximation, TEM01 fields excited by angular misalignments can be expressed as
\begin{align}
    \vec{E}_{01} &\approx (I-\hat{M}_{01})^{-1}\hat{\Theta}(I-\hat{M}_{00})^{-1}\vec{E}_{00,\mathrm{in}}. \label{approx}
\end{align}
%\color{red} the statement below may be too tough to understand for the readers.\color{black}
One can express TEM10 fields introduced by angular misalignments in the same manner.

%\begin{figure}[t]
%\centering
%\includegraphics[width=250pt]{block_diagram_FP.eps}
%\caption{The block diagram picture of Fabry-Perot cavity with angular misalignment. $\hat{P}$ denotes the propagator between two mirrors, and each $\hat{r}$ and $\hat{t}$ represent reflection and transmission matrices, respectively. Angular misalignments are injected as perturbations at each node, 1 and 2.\color{red} perturbations have to be drawn with some kind of summing node\color{black}}
%\label{block_diagram_FP}
%\end{figure}

%This method can be applied for more complex optical system.
%We can define the field of TEM00, TEM10 and TEM01 at each node for any optical system.
%In addition, we can define the matrix, $\hat{\Theta}$, which introduce excited modes at each node due to angular misalignments.
%The fields of excited higher order modes, TEM10 and TEM01, are also expressed as Eq. (\ref{E01}).
%Assuming misalignment angle is small, the excited fields can be approximated as Eq. (\ref{approx}).
%Hence, the linear approximation method we presented here can be applied to any optical systems.

Our method, in comparison to the previous study~\cite{Hefetz1997}, obviously simplifies the resulting solutions at a cost of losing the accuracy in particular when the amount of misalignment is large. However, in practical experiments, the angular drifts are suppressed by active controls and therefore the amount of misalignment is typically small enough that our approximation is valid.
%Therefore, our linear approximation method simplifies the computation while it is still useful for experiments.

\begin{figure}[t]
\centering
  \includegraphics[width=\linewidth]{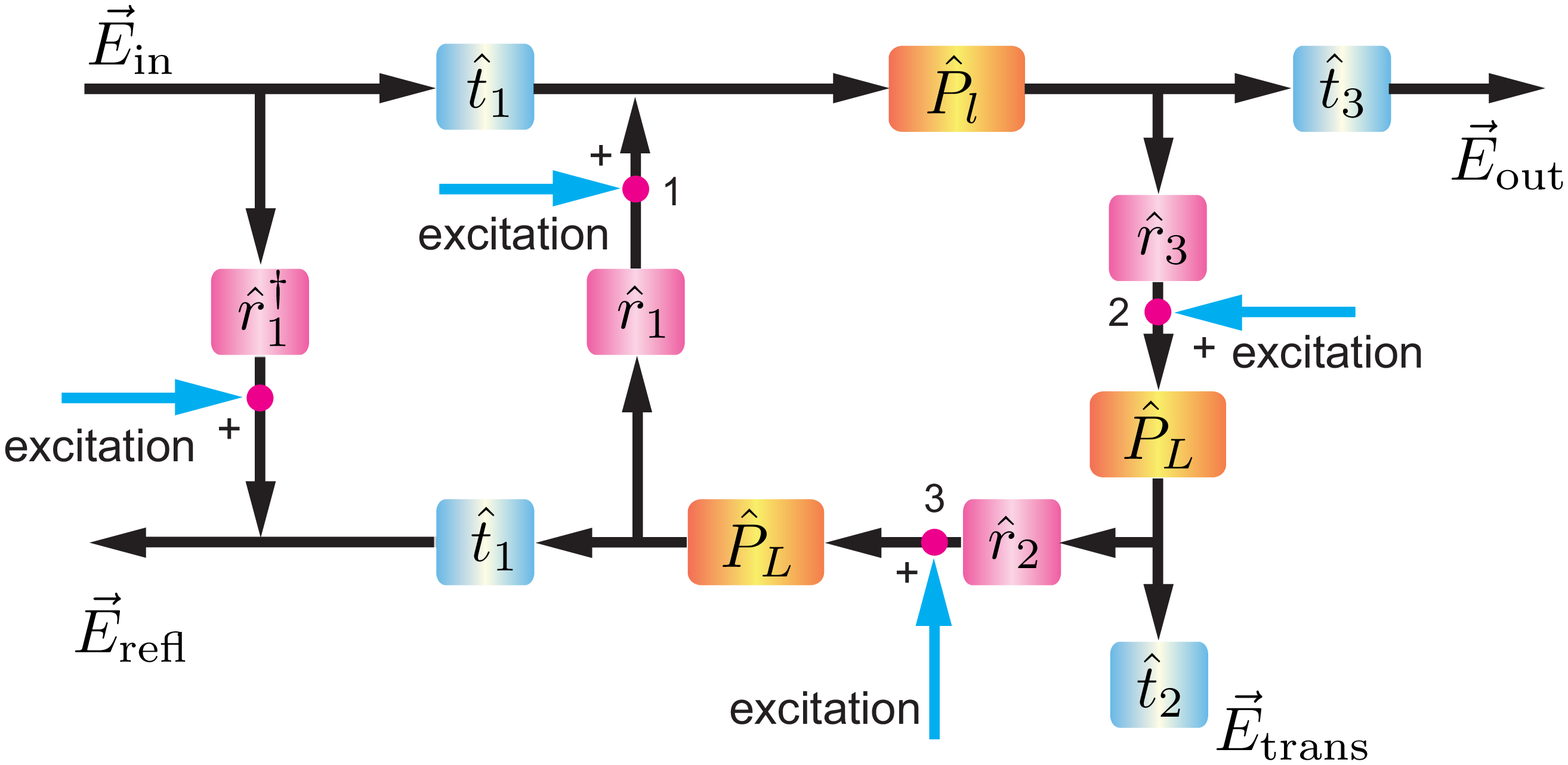}
\caption{Block diagram description of a triangular cavity with angular misalignments. $\hat{P}$ denotes the free-space propagator between two mirrors and its subscript, $L$ or $l$, correspond to the length of propagation. Each $\hat{r}$ and $\hat{t}$ represent reflection and transmission matrices, respectively. The labled number, 1 to 3, denote the nodes. Angular misalignments are injected as perturbations at each node. It should be noted that promptly reflected light with MC1 angular misalignment couples to the WFS signals.}
\label{block_triangular}
\end{figure}

Besides, another advantage of the linearization is that the entire analysis can be expressed by a block diagram as shown in Fig. \ref{block_triangular} similarly to those used in the classical feedback control theory~\cite{lurie2000classical}. Such block diagrams can help ones to acquire comprehensive physical pictures.

%A linear approximation method which we presented is a comprehensive and useful approach for analysis of optical systems.

\section{Angular Response of a Triangular Cavity}
%So far, the mathematical foundations to compute the angular response have been presented.
%In this section, we will apply the linear approximation method to a triangular cavity in order to understand its angular response.
We now consider the case in which a triangular cavity is in the form of isosceles triangle with the length of one side much shorter than the other two as shown in Fig. \ref{IMC_top}.
The incoming laser beam is assumed to be injected to the shorter side.
Fig. \ref{block_triangular} shows a block diagram for the system with the mirrors misaligned by some small amounts.
%By using this diagram, the impact of angular misalignments can be computed by diagonal matrices.
%Hence, the calculation process of angular response becomes quite simple. 

The free-space propagator can be defined as
\begin{align}
\hat{P}_{L,l} &= 
\mathrm{diag}\left(\mathrm{e}^{-i\Phi_{L,l}}, \mathrm{e}^{-i(\Phi_{L,l}-\eta_{L,l})}, \mathrm{e}^{-i(\Phi_{L,l}-\eta_{L,l})}\right),
\end{align}
where its subscript, $L$ and $l$, correspond to the length of propagation.
The matrices $\hat{t}$ and $\hat{r}$ denote the transmission and reflection matrix of the mirror as,
\begin{align}
\hat{r}_{i} &=
\mathrm{diag}(r_i, -r_i, r_i),\\
\hat{t}_{i} &=
\mathrm{diag}(t_i, t_i, t_i),
\end{align}
where the subscripts, $i=1,2,3$, express each mirror of the IMC.
It should be noted that the sign of reflectivity for TEM10 is negative due to the mirroring effect which flips the coordinates left-to-right in the horizontal plane.
Combining these matrices, the cavity round-trip propagator, $\hat{P}_{\mathrm{rt}}$, can be expressed as
\begin{align}
\hat{P}_{\mathrm{rt}} &\equiv \hat{P}_L \hat{r}_2 \hat{P}_L \hat{r}_3  \hat{P}_l \hat{r}_1 \notag \\
 &= r  \mathrm{e}^{-i\Phi_{\mathrm{rt}}}
\mathrm{diag}\left(1, -\mathrm{e}^{i\eta_{\mathrm{rt}}}, \mathrm{e}^{i\eta_{\mathrm{rt}}}\right),
\end{align}
where $r\equiv r_1r_2r_3$ is a total reflectivity of the system which is merely a multiplication of the amplitude reflectivity from all the mirrors and $\eta_{\mathrm{rt}}$ represents a round-trip Gouy phase.
Using the round-trip propagator, one can express the fields inside the cavity as
\begin{align}
    \vec{E}_{\mathrm{cav}} &= (I-\hat{P}_{\mathrm{rt}})^{-1}\hat{t}_1\vec{E}_{\mathrm{in}}.
    \label{Ecav}
\end{align}

In practical experiments, the phase modulated laser beam which can be split into the carrier and the sidebands is widely used in order to lock the laser to the cavities as shown in Fig. \ref{IMC_top}.
Assuming that the cavity is kept resonant for the carrier i.e., $\mathrm{e}^{-i\Phi_{\mathrm{rt}}}=1$, then one can define the cavity gain for the carrier in the form of each HG modes as 
\begin{align}
G_{00} &\equiv \frac{1}{1-r}, \\
G_{10} &\equiv \frac{1}{1+r\mathrm{e}^{i\eta_{\mathrm{rt}}}}, \\
G_{01} &\equiv \frac{1}{1-r\mathrm{e}^{i\eta_{\mathrm{rt}}}}.
\end{align}
It should be noted that the sign of cavity gain is flipped between TEM01 and TEM10 modes due to the mirroring effect for horizontal axis.

%\color{red} Too sudden to talk about sidebands. Need some mention beforehand\color{black}
On the other hand, the cavity gain for the sidebands which are kept off-resonant can be written as
\begin{align}
G_{00}^{\mathrm{SB}} &\equiv \frac{1}{1+r}.
\end{align}
Here we assumed that the reflectivity for both upper and lower sidebands are identical.
When the reflectivities of mirrors are high i.e., $r_i\approx1$, the cavity gain for the sidebands are much smaller than that of the carrier.
Therefore, ignoring the effect of the sidebands inside the cavity, we will consider only the carrier field for the intra-cavity fields.
Then Eq. (\ref{Ecav}) can be rewritten as
\begin{align}
    \vec{E}_{\mathrm{cav}} &= \mathrm{diag}(G_{00}, G_{10}, G_{01})\hat{t}_1\vec{E}_{\mathrm{in}}.
\end{align}
Thus each HG mode excited by angular misalignment is built up inside the cavity as characterized by each cavity gain.

%Assuming the incoming beam is pure TEM00 mode, one can see that all higher-order-modes which are existing inside the cavity are generated by angular misalignments.
%When angular misalignments exist, the built up TEM00 mode couples into TEM10 or TEM01 modes and they are built up again.
%Here we treat the excited higher-order-modes as perturbations as shown in \zu{block_triangular} and take the case of MC3 pitch misalignment.
%Let us compute the response of the system when the MC3 mirror is misaligned in pitch. The perturbation at MC3 can be written as 
%\begin{align}
    %\Theta_{\mathrm{MC3}} &= -i\Theta_3 \times r_3\mathrm{e}^{-i\Phi_{2l}}G_{00}\la 00|E \ra_{\mathrm{in}}.
%\end{align}
%\color{red}???\color{black}After building up inside the cavity, the excited field at excited node, node 2, becomes
%\begin{align}
    %\la 01|E \ra_{\mathrm{MC3}}&= -i\Theta_3 \times r_3\mathrm{e}^{-i\Phi_{2l}}G_{00}G_{01}\la 00|E %\ra_{\mathrm{in}}.
%\end{align}
%\color{red} Sounds like this is a repeat of what you have shown in the previous section with a single FP. Nencessary?\color{black}
%By multiplying appropriate propagators, one can obtain the fields of excited higher-order modes at each node.
%We can conduct calculations for other mirrors and another higher-order mode as the same manner.
%In this way, we can compute angular response of a triangular cavity by simple calculation.

\subsection{Beam Spot Response}
\begin{figure}[t]
\centering
\includegraphics[width=250pt]{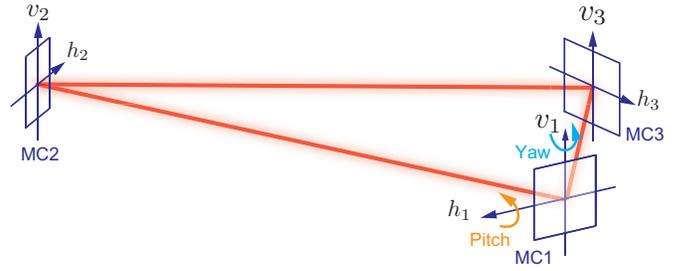}
\caption{Beam spot coordinates on mirrors and the definition of pitch and yaw rotations. Pitch and yaw are rotations around horizontal and vertical axes on a mirror, respectively.}
\label{coordinate}
\end{figure}
The response to the beam spot positions is one of the most important properties to understand the behavior of a triangular cavity.
%Here we will calculate how the beam spot positions will change due to angular misalignments.
The lateral shift of the beam on each mirror can be derived by evaluating the real part of the coefficients for the 01 and 10 modes. For the purpose, let us expand a field into real and imaginary parts as,
\begin{align}
A\left(U_{00} + CU_{01,10}\right)
\end{align}
where $C$ is a complex number which can be expressed using a small beam translation $a$ and a small tilt $\alpha$ \cite{Kawabe1994},
\begin{align}
C &= \frac{a}{w_0} + i\frac{kw_0}{2}\alpha, \label{beam_shift}
\end{align}
where $w_0$ is the beam size at its waist and $k$ is the wavenumber.
\shiki{beam_shift} tells the real part of $C$ denotes the beam translation and the imaginary part denotes the beam tilt.
Therefore, the beam spot translation can be derived from the real part of normalized coefficient, $C$, of TEM01/10.
The notations of angular tilts and axes on the mirrors are as shown in Fig. \ref{coordinate}.

%high ref. approximation
%When each mirror has an angular tilt, $\theta_i$, one can obtain beam spot change by calculating real part of complex coefficient $C$ at each mirror position.
Evalating the real part of the field on each mirror and approximating all the mirrors to be a perfect reflector i.e., $r_i=1\,(i=1,2,3)$, one can find the following relation between the mirror tilts and resulting variation in the beam spot as
\begin{align}
\begin{pmatrix}
y_1 \\
y_2 \\
y_3 \\
\end{pmatrix}
&= -R
\begin{pmatrix}
g & 1 & g+\frac{2l}{R} \\
1 & 1 & 1 \\
g+\frac{2l}{R} & 1 & g \\
\end{pmatrix}
\begin{pmatrix}
\theta_1 \\
\theta_2 \\
\theta_3 \\
\end{pmatrix},
\label{pit_beam}
\end{align}
where $g$ is the g-factor, defined as
\begin{align}
g &\equiv 1 - \frac{L+l}{R},
\end{align}
and $y_i \, (i=1,2,3)$ describe vertical displacements of the beam along the $y$-axis.
%We have used the following relationships as well,\color{red} Not sure if we need to mention Zr and w0 at this point... Put them in appexndix?\color{black}
%\begin{align}
%z_{\mathrm{R}} &= \sqrt{(L+l)(R-L-l)}, \\
%w_0 &= \sqrt{\frac{z_{\mathrm{R}}\lambda}{\pi}},
%\end{align}
%where $z_{\mathrm{R}}$ represents the Rayleigh length and $w_0$ represents the beam size at its waist.
A detailed calculation for obtaining Eq. (\ref{pit_beam}) is presented in Appendix A.

Since MCs 1 and 3 have an incident angle of approximately $\pi/4$, the $y_1$ axis and the $y_3$ axis are rotated $\pi/4$ around the $z$ axis with respect to the $y$ axis \cite{Freise2013Finesse}.
Therefore, an effective misalignment angle in pitch is reduced by $\cos(\pi/4) = 1/\sqrt{2}$, so that
%In addition, $\psi$ has opposite sign, and be expressed as
\begin{align}
\psi_{1,3} &= \frac{1}{\sqrt{2}}\theta_{1,3}.
\end{align}
The axis $y$ is same as the axis on the mirror $v$.
Hence the beam spot displacement on the mirror can be written as
\begin{align}
\begin{pmatrix}
v_1 \\
v_2 \\
v_3 \\
\end{pmatrix}
&= -\frac{R}{\sqrt{2}}
\begin{pmatrix}
g & \sqrt{2} & g+\frac{2l}{R} \\
1 & \sqrt{2} & 1 \\
g+\frac{2l}{R} & \sqrt{2} & g \\
\end{pmatrix}
\begin{pmatrix}
\psi_1 \\
\psi_2 \\
\psi_3 \\
\end{pmatrix}.
\label{matrix1}
\end{align}
\shiki{matrix1} shows us how the eigenmode of triangular cavity transforms as a function of the tilt of each mirror.

Mirror positions may not always be preferable DoFs. They can be converted into the ones representing the translation and tilt of the beam at the waist, and the spot position on MC2 as
\begin{align}
\begin{pmatrix}
y_{\mathrm{w}} \\
\theta_{\mathrm{w}} \\
v_2 \\
\end{pmatrix}
\equiv
\begin{pmatrix}
\frac{v_1+v_3}{2} \\
\frac{v_1-v_3}{2l} \\
v_2 \\
\end{pmatrix}
=
\begin{pmatrix}
\frac{1}{2} & 0 & \frac{1}{2} \\
\frac{1}{2l} & 0 & -\frac{1}{2l} \\
0 & 1 & 0 \\
\end{pmatrix}
\begin{pmatrix}
v_1 \\
v_2 \\
v_3 \\
\end{pmatrix}.
\end{align}
In addition, we convert the actuation basis into MCs 1 and 3 common and differential motions expressed as 
\begin{align}
\begin{pmatrix}
\psi_{\mathrm{c}} \\
\psi_{\mathrm{d}} \\
\psi_2 \\
\end{pmatrix}
\equiv
\begin{pmatrix}
\frac{\psi_1+\psi_3}{2} \\
\frac{\psi_1-\psi_3}{2} \\
\psi_2 \\
\end{pmatrix}
=
\begin{pmatrix}
\frac{1}{2} & 0 & \frac{1}{2} \\
\frac{1}{2} & 0 & -\frac{1}{2} \\
0 & 1 & 0 \\
\end{pmatrix}
\begin{pmatrix}
\psi_1 \\
\psi_2 \\
\psi_3 \\
\end{pmatrix}.
\end{align}
By using these bases, the beam spot response to pitch angular misalignment can be summarized as
\begin{align}
\begin{pmatrix}
y_{\mathrm{w}} \\
\theta_{\mathrm{w}} \\
v_2 \\
\end{pmatrix}
&=
\begin{pmatrix}
\sqrt{2}(L-R) & 0 & -R \\
0 & \sqrt{2} & 0 \\
-\sqrt{2}R & 0 &-R \\
\end{pmatrix}
\begin{pmatrix}
\psi_{\mathrm{c}} \\
\psi_{\mathrm{d}} \\
\psi_2 \\
\end{pmatrix}.
\label{pitch_waist}
\end{align}
\shiki{pitch_waist} shows that misalignment in MC2 cannot access to the tilt at the waist position, and only shift the plane of cavity eigenmode. 
Common angular tilt rotates cavity eignemode around the center of curvature as shown in figure 19. in \cite{Kawazoe2011}.
Differential tilt does not change the spot on the curved mirror and the waist, though the tilt at waist is rotated.

In the same way, yaw misalignment can be computed as
\begin{align}
\begin{pmatrix}
x_1 \\
x_2 \\
x_3 \\
\end{pmatrix}
&=
\begin{pmatrix}
\xi & -\frac{l}{g} & -\xi \\
-\frac{l}{g} & \frac{L+l}{g} & -\frac{l}{g} \\
-\xi & -\frac{l}{g} & \xi \\
\end{pmatrix}
\begin{pmatrix}
\phi_1 \\
\phi_2 \\
\phi_3 \\
\end{pmatrix}.
\label{yaw_beam}
\end{align}
where $xi$ is defined as $\xi\equiv L+l(1-L/R)/g$.
When the displacements perpendicular to the beam, $x_1$ and $x_3$, are projected on each mirror, MCs 1 and 3, the displacements on the mirror surface are enhanced by $1/\cos(\pi/4) = \sqrt{2}$.
Therefore, Eq. (\ref{yaw_beam}) can be written as
\begin{align}
\begin{pmatrix}
h_1 \\
h_2 \\
h_3 \\
\end{pmatrix}
&=
\begin{pmatrix}
\sqrt{2}\xi & -\sqrt{2}\frac{l}{g} & -\sqrt{2}\xi \\
-\frac{l}{g} & \frac{L+l}{g} & -\frac{l}{g} \\
-\sqrt{2}\xi & -\sqrt{2}\frac{l}{g} & \sqrt{2}\xi \\
\end{pmatrix}
\begin{pmatrix}
\phi_1 \\
\phi_2 \\
\phi_3 \\
\end{pmatrix}.
\end{align}
The bases can be converted in the manner same as that for pitch by using the relationship expressed as
\begin{align}
&\begin{pmatrix}
x_{\mathrm{w}} \\
\theta_{\mathrm{w}} \\
h_2 \\
\end{pmatrix}
\equiv
\begin{pmatrix}
\frac{(h_1+h_3)\cos(\pi/4)}{2} \\
\frac{(h_1-h_3)\cos(\pi/4)}{2l} \\
h_2 \\
\end{pmatrix}
=
\begin{pmatrix}
\frac{1}{2\sqrt{2}} & 0 & \frac{1}{2\sqrt{2}} \\
\frac{1}{2\sqrt{2}l} & 0 & -\frac{1}{2\sqrt{2}l} \\
0 & 1 & 0 \\
\end{pmatrix}
\begin{pmatrix}
h_1 \\
h_2 \\
h_3 \\
\end{pmatrix}, \\
&\begin{pmatrix}
\phi_{\mathrm{c}} \\
\phi_{\mathrm{d}} \\
\phi_2 \\
\end{pmatrix}
\equiv
\begin{pmatrix}
\frac{\phi_1+\phi_3}{2} \\
\frac{\phi_1-\phi_3}{2} \\
\phi_2 \\
\end{pmatrix}
=
\begin{pmatrix}
\frac{1}{2} & 0 & \frac{1}{2} \\
\frac{1}{2} & 0 & -\frac{1}{2} \\
0 & 1 & 0 \\
\end{pmatrix}
\begin{pmatrix}
\phi_1 \\
\phi_2 \\
\phi_3 \\
\end{pmatrix}.
\end{align}
It should be noted that the effect of $\sqrt{2}$ enhancement due to the projection on MCs 1 and 3 is cancelled.
Then the beam spot response to yaw misalignments can be expressed as
\begin{align}
\begin{pmatrix}
x_{\mathrm{w}} \\
\theta_{\mathrm{w}} \\
h_2 \\
\end{pmatrix}
&=
\begin{pmatrix}
0 & -2L & 0 \\
\frac{2(-L+R)}{L-R+l} & 0 & -\frac{R}{L-R+l} \\
\frac{2Rl}{L-R+l} & 0 & -\frac{R(L+l)}{L-R+l} \\
\end{pmatrix}
\begin{pmatrix}
\phi_{\mathrm{c}} \\
\phi_{\mathrm{d}} \\
\phi_2 \\
\end{pmatrix}.
\label{yaw_waist}
\end{align}
Eq. (\ref{yaw_waist}) shows that only misalignment in MCs 1 and 3 differential motion can access to the beam waist shift.
The beam waist tilt and MC2 spot position change can be expressed by a combination of MCs 1 and 3 differential motion and MC2 motion.

These results are fully consistent with the previous analysis based on the geometrical argument~\cite{Kawazoe2011}. %though there is a difference due to the definition of MCs 1 and 3 common and differential motion.
Therefore, this linear approximation offers an alternative method to compute variations in the spot positions without elaborating the geometrical analysis.

\subsection{Wave Front Sensing Signals of the IMC}
%A triangular cavity is used as a mode cleaning cavity in gravitational wave detectors which is called IMC \cite{Mueller2016}.
%Maintaining the alignment of IMC is crucial in order to achieve the detector operation with the best performance.
%Therefore, ASC using WFS signals are adopted to IMC for the reduction of angular misalignment.
%In order to design ASC, it is necessary to understand the behavior of WFS signals.
%We will calculate WFS signals of IMC using the linear approximation method which presented previous section.
%First, we will briefly introduce the WFS signals, then compute WFS signals.
The WFS technique is sensitive to small tilts of the mirrors.
The signals are obtained by demodulating the intensity information detected by quadrant photodiodes (QPDs) \cite{Hefetz1997} in similar manner to the Pound-Drever-Hall (PDH) technique \cite{Drever1983}.
When one of the mirrors is misaligned, it excites the second lowest-order transverse modes, namely the TEM10 or TEM01 in the carrier.
The WFS signals are essentially the interference between the TEM00 of the sidebands and the TEM01 or TEM10 modes of the carrier.

We assume that the incoming light is a purely phase modulated TEM00 mode described as
\begin{align}
\vec{E}_{\mathrm{in}} &=
\begin{pmatrix}
\la00|E\ra_{\mathrm{in}} \\ 0 \\ 0
\end{pmatrix}.
\end{align}
Phase modulated input laser beam, which is used in PDH scheme, can be expressed as
\begin{align}
\la 00|E\ra_{\mathrm{in}} \approx \left[J_0(m)\mathrm{e}^{i\Omega t} + J_1(m)\mathrm{e}^{i(\Omega+\omega_m)t} - J_1(m)\mathrm{e}^{i(\Omega-\omega_m)t}\right] \notag \\
    \times \la 00|E \ra
\end{align}
where $\Omega$ corresponds to the angular frequency of the carrier field,  $\omega_{\mathrm{m}}$ is the modulation angular frequency, $m$ is the modulation depth and $J_{k}(m)$ is Bessel functions of the first kind.

By measuring the differential of reflected field using a horizontally and vertically split photodiodes, QPD, for pitch and yaw, one can obtain the intensity information, $S_{\mathrm{QPD}}$, which can be expressed as
\begin{equation}
S_{\mathrm{QPD}} \propto \int_0^{\infty}\mathrm{d}x \left(\vec{E}_{\mathrm{refl}}^{\dagger}\cdot \vec{E}_{\mathrm{refl}}\right) - \int^0_{-\infty}\mathrm{d}x \left(\vec{E}_{\mathrm{refl}}^{\dagger} \cdot\vec{E}_{\mathrm{refl}}\right).
\end{equation}
The fields at reflection port of each mode can be calculated by multiplying propagators.
The WFS signals are obtained by demodulating the signals, $S_{\mathrm{QPD}}$, with the modulation angular frequency, $\omega_{\mathrm{m}}$.
After some math, one can derive the WFS signals described as
\begin{align}
\mathrm{WFS} = iP_0J_0(m)J_1(m)\big[&U_{00}^* (G_{00}G_{10,01}\mathrm{e}^{i\bar{\eta}}\Theta U_{10,01}) \\ 
&- U_{00}(G_{00}G_{10,01}\mathrm{e}^{i\bar{\eta}}\Theta U_{10,01})^*\big],
\end{align}
where $P_0$ denotes the optical power of the input laser beam and $\bar{\eta}$ is a spcific Gouy phase which depends on the misaligned mirror.
When we apply parameters of KAGRA IMC, the WFS signals can be calculated as shown in \zu{KAGRA_yaw}.

On the other hand, aLIGO adopted a different geometry in which the laser beam is injected into the longer side of the IMC.
%Although the beam propagating direction is different from shorter-side-injection geometry, one can adopt the same calculation method and obtain WFS signals.
\zu{LIGO} shows the WFS signals of aLIGO geometry IMC.
Since MCs 1 and 3 WFS signals are highly degenerate, it is hard to distinguish between them.

\begin{figure}[t]
\centering
\includegraphics[width=\linewidth]{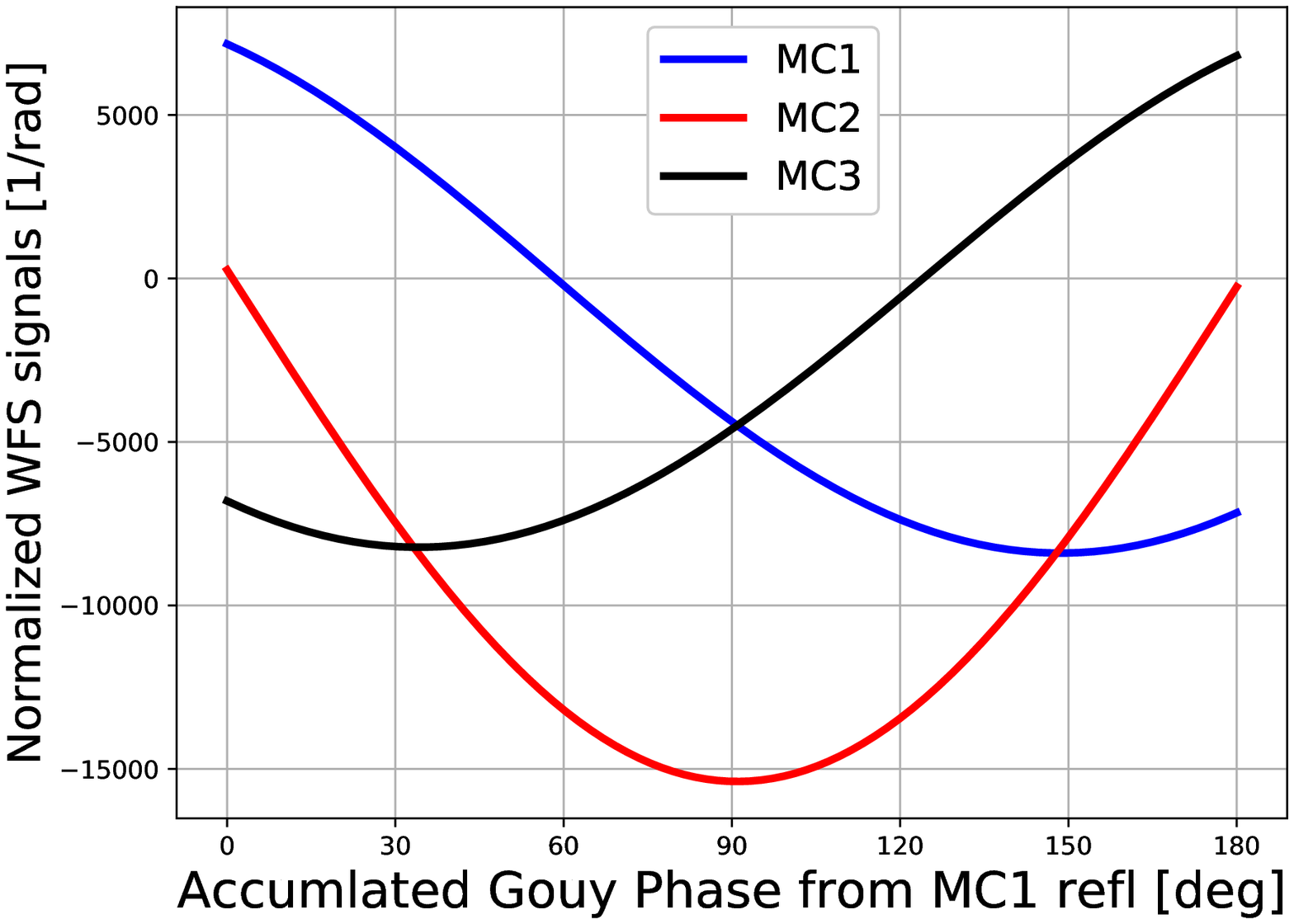} \\
\includegraphics[width=\linewidth]{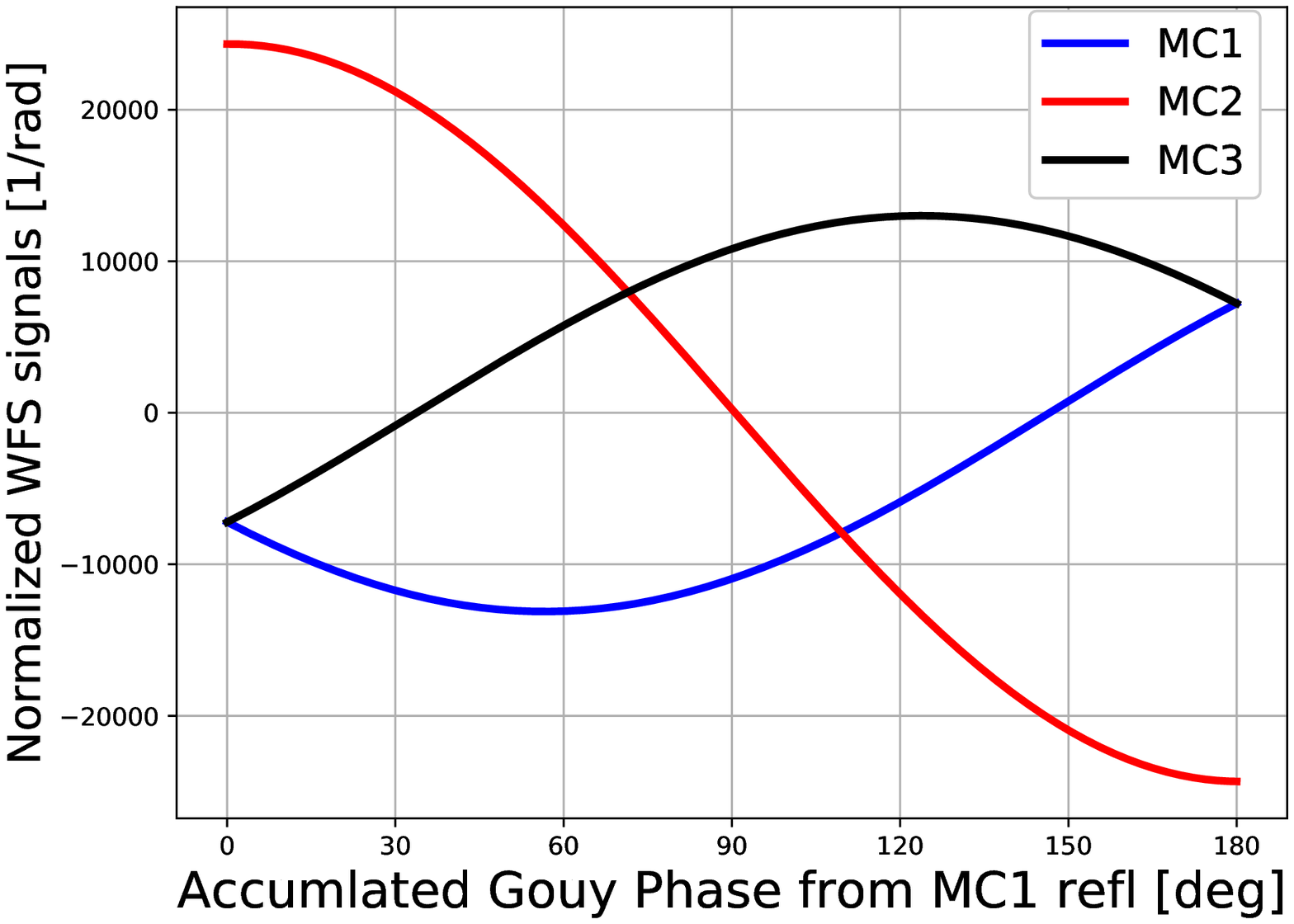}
\caption{Pitch (top) and yaw (bottom) WFS signals of shorter-side-injection geometry the IMC. We assumed that the geometrical parameters of the IMC are same as KAGRA.}
\label{KAGRA_yaw}
\end{figure}

\begin{figure}[t]
\centering
\includegraphics[width=\linewidth]{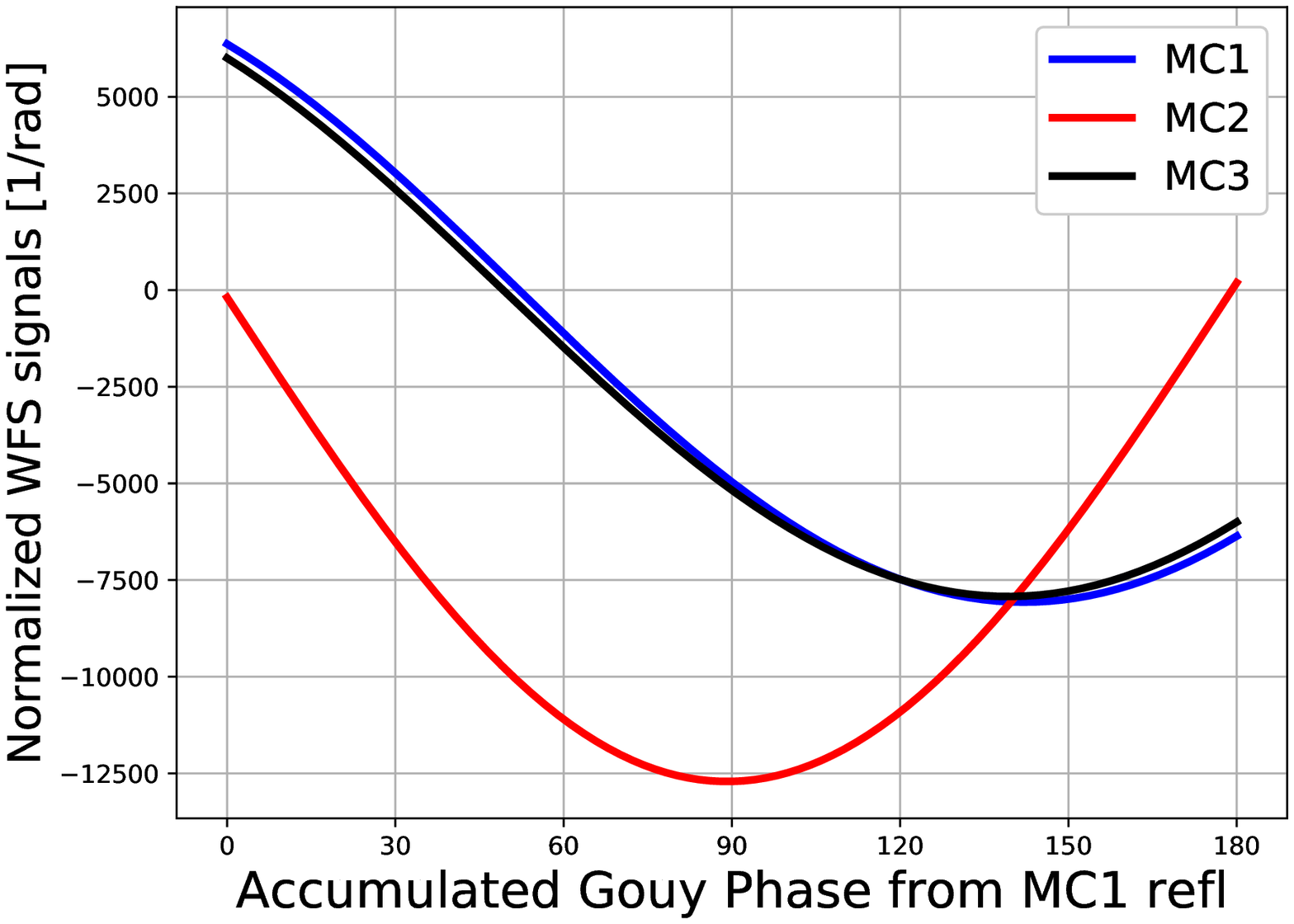} \\
\includegraphics[width=\linewidth]{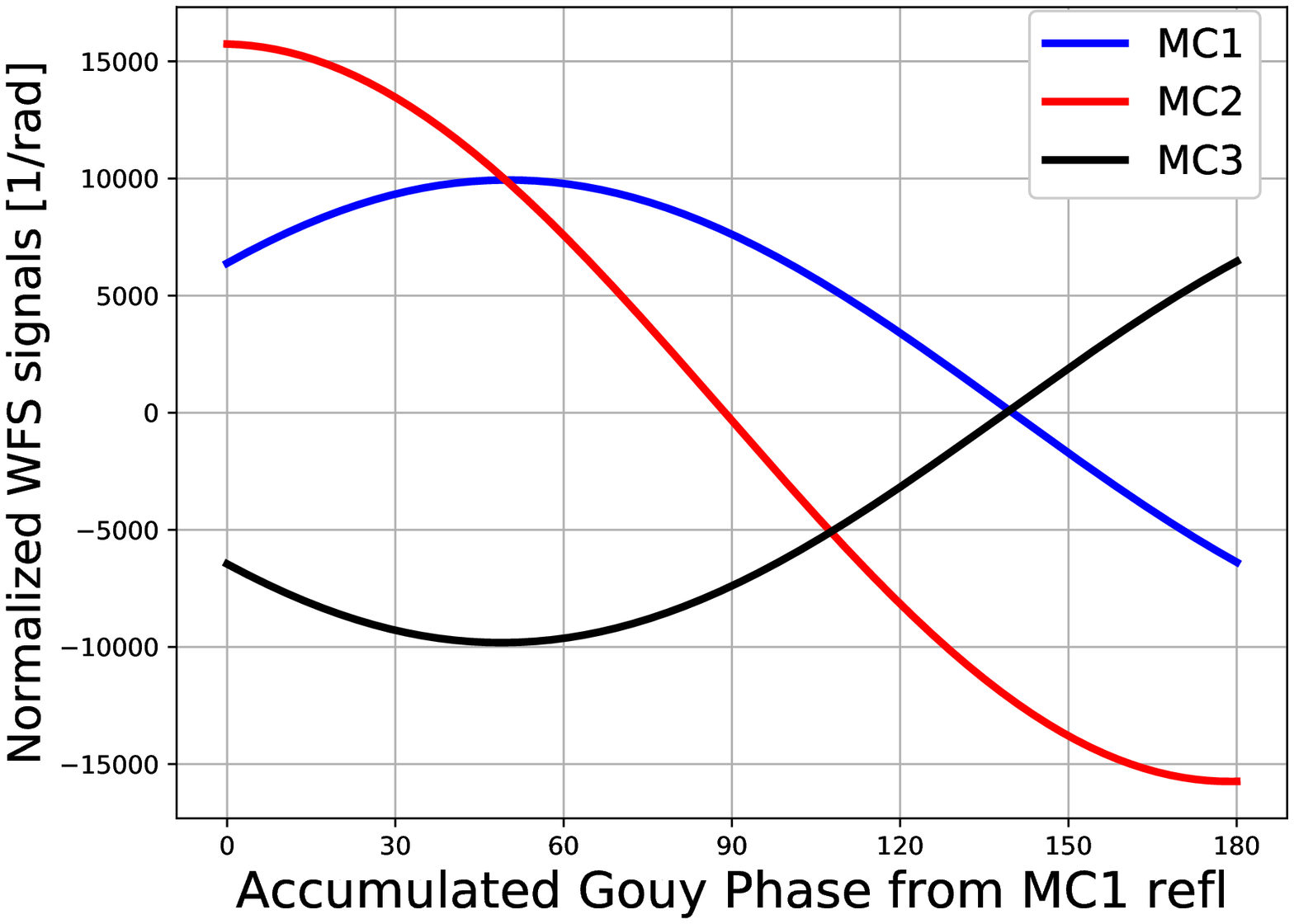}
\caption{Pitch (top) and yaw (bottom) WFS signals of longer-side-injection geometry. We assumed that the geometrical parameters of the IMC are same as aLIGO.}
\label{LIGO}
\end{figure}

\section{Implication to Experiment}
%We have calculated WFS signals of IMCs used in gravitational wave detectors by the linear approximation method.
The IMC has three DoFs for angular misalignments in both pitch and yaw. On the other hand, the number of observable DoFs by the WFS technique is only two since the WFSs sense the difference between the two beams --- the incoming and intra-cavity beams.
In this section, the observable degrees of freedom (DoFs) of the IMC are discussed for the case where the incoming beam is injected to the shorter side of the IMC such as KAGRA \cite{Aso2013}.
For the other case where the incoming beam is injected to longer side of the IMC such as aLIGO \cite{Mueller2016}, the results are shown in Appendix B for completeness.

The singular value decomposition (SVD) is one of the useful tools in order to understand what DoFs can or cannot be sensed. The SVD is widely used for data analysis \cite{Alter2000}.
%Similarly to SVD method, we can decompose the sensing matrices and obtain orthogonal bases which indicate the observable DoFs of IMC WFS.
The WFS signals contain beam tilt and translation components and are expressed as bellow by using a sensing matrix $\hat{S}$,
\begin{equation}
    \begin{pmatrix}
    \mathrm{tilt} \\
    \mathrm{trans.} \\
    \end{pmatrix}
=
\hat{S}
    \begin{pmatrix}
    \theta_1 \\
    \theta_2 \\
    \theta_3 \\
    \end{pmatrix},
\end{equation}
where $\hat{S}$ is a $2\times3$ matrix and its elements can be written by using the cavity gain and Gouy phase shift.
This sensing matrix can be expressed by geometrical parameters with high reflective approximation ($r_i\approx1$) and incoming beam is pure TEM00 mode.
Then we will apply SVD to rewrite them to determine which DoFs are observable by the WFS technique.

In the case of the geometry used in KAGRA \cite{Aso2013}, one can obtain the relationship between the WFS signals and mirror tilt angles in pitch as
\begin{align}
\begin{pmatrix}
	\mathrm{tilt} \\
	\mathrm{trans.}\\
\end{pmatrix}_{\mathrm{Pitch}}
&= \sqrt{2}\gamma w_0
\begin{pmatrix}
	g\gamma & 0 & -g\gamma \\
	-g & -\sqrt{2} & -g \\
\end{pmatrix}
\begin{pmatrix}
	\psi_1 \\
	\psi_2 \\
	\psi_3
\end{pmatrix},
\end{align}
where $\gamma\equiv(L+l)/z_{\mathrm{R}}$, and $w_0$ is beam size at its waist.
%Here we used geometrical parameters to express sensing matrix.
Adopting the SVD analysis, one can arrive at
\begin{align}
\hat{S}_{\mathrm{Pitch}} &=
\sqrt{2}\gamma w_0I
\begin{pmatrix}
	g\gamma & 0 & 0 \\
	0 & 1 & 0
\end{pmatrix}
\begin{pmatrix}
	1 & 0 & -1 \\
	-g & -\sqrt{2} & -g \\
	1 & -\sqrt{2}g & 1
\end{pmatrix}.
\end{align}
It is clear that one the DoF shown in the lowest row in the rightmost matrix.
On the other hand, the first and second rows are observable. They correspond to MCs 1 and 3 differential motion and the combination of MCs 1 and 3 common and MC2 motion, respectively.
%１行目はdiff.に対応し、２行目はコモンとMC2のモーションの組み合わせに対応する
Comparing to \shiki{pitch_waist}, one can identify that the observable DoFs correspond to the beam waist tilt and translation.

For the case of yaw, the sensing matrix can be written as
\begin{align}
\begin{pmatrix}
	\mathrm{tilt} \\
	\mathrm{trans.}\\
\end{pmatrix}_{\mathrm{Yaw}}
&= \sqrt{2}\gamma w_0
\begin{pmatrix}
	-g & 1 & -g \\
	-g\gamma & 0 & g\gamma
\end{pmatrix}
\begin{pmatrix}
	\phi_1 \\
	\phi_2 \\
	\phi_3
\end{pmatrix}.
\end{align}
The decomposed form can be expressed as
\begin{align}
\hat{S}_{\mathrm{Yaw}} &= 
\sqrt{2}\gamma w_0I
\begin{pmatrix}
	g & 0 & 0 \\
	0 & g\gamma & 0
\end{pmatrix}
\begin{pmatrix}
	-1 & \frac{1}{g} & -1 \\
	-1 & 0 & 1 \\
	1 & 2g & 1
\end{pmatrix}.
\end{align}
The observable DoFs are the combination of MCs 1 and 3 common and MC2 motion and MCs 1 and 3 differential motion.
Similarly to the pitch case, these observable DoFs correspond to the beam waist tilt and translation.
Therefore, when the beam is injected to the shorter side of the IMC, the beam waist tilt and translation are observable. %which correspond to the beam axis injected to the main interferometer.

\section{Conclusion}
We have presented the linear approximation method that can simplify the analytical calculation of angular response of optical cavities.
The method enables us not only to simplify the calculation but also to make equivalent block diagrams which can bring comprehensive pictures.

The method was applied to a triangular cavity called the input mode cleaning cavity.
%Especially, an alignment sensing and control system using WFS signals is indispensable to achieve the stabilized input beam for main interferometer.
%The linear approximation approach makes it easier to calculate the angular response of optical cavities with the assumption of incoming beam is well aligned.
The angular response to the shift in the beam spots and the WFS signals of the IMC in gravitational wave detectors were explicitly given by this approach.
The results are consistent with the previous report which conducted a
completely different approach --- geometrical analysis.

%It should be noted that we assumed the input beam is well aligned and has purely phase modulated TEM00 mode.
%In practice, the input beam has beam jitters i.e., TEM01,10 modes, and they couple to angular response of the fields.
%It is difficult, however, to take into consideration input beam misalignments due to their Gouy phase dependence which cannot be estimated precisely.
%Therefore, the actual behaviors of WFS signals may deviate a little from the obtained result by this approach.

Finally, the WFS sensing matrices are analyzed via the singular value decomposision scheme in order to study the observable DoFs of the IMC.
When the incoming beam is injected to the shorter side of the IMC, the observable DoFs by the WFS scheme are determined to be the beam waist tilt and translation.

The linear approximation method presented in this article can be applied to optical systems which are more involved, including the main interferometer of gravitational wave detectors. 
The linear approximation method should be able to give deeper insight into the fundamental comprehension of the gravitational wave interferometers. %alignment control and the observable DoFs by combining with SVD method.

\appendix
\section{Calculation of Beam Spot Motion}
In this section, we will show one example of angular response calculation --- MC3 pitch misalignment.
TEM01 field at each node which is excited by MC3 pitch misalignment can be computed by multiplying propagators.
The cavity gain for TEM01 can be written as
\begin{align}
    G_{01} &= \frac{1-\cos\eta_{\mathrm{rt}}+i\sin\eta_{\mathrm{rt}}}{2(1-\cos\eta_{\mathrm{rt}})}.
\end{align}
Here we introduce high reflective approximation i.e., assuming $r\approx1$.
In order to calculate the normalized coefficient, $C$, propagators must be taken into account.
Then the normalized coefficient, $C$, at each node $1$ to $3$ can be written as
\begin{align}
    C_1 &= -i\Theta_3\frac{\mathrm{e}^{i\eta_{\mathrm{MC3\to MC1}}}-\mathrm{e}^{i(\eta_{\mathrm{rt}}-\eta_{\mathrm{MC3\to MC1}}})}{2(1-\cos\eta_{\mathrm{rt}})} \\
    C_2 &= -i\Theta_3\frac{1-\cos\eta_{\mathrm{rt}}+i\sin\eta_{\mathrm{rt}}}{2(1-\cos\eta_{\mathrm{rt}})} \\
    C_3 &= -i\Theta_3\frac{\mathrm{e}^{i\eta_{\mathrm{MC3\to MC2}}}-\mathrm{e}^{i(\eta_{\mathrm{rt}}-\eta_{\mathrm{MC3\to MC2}}})}{2(1-\cos\eta_{\mathrm{rt}})}.
\end{align}
where the subscripts of Gouy phase shift such as $\mathrm{MC3\to MC1}$ denote the propagation of the beam.
Since the beam waist locates on halfway between MCs 1 and 3 and the length of shorter side, $2l$, is much smaller than that of longer side,$L$, Gouy phase shift can be computed as
\begin{align}
    \eta_{\mathrm{MC3\to MC2}} &= \arctan\left(\frac{L+l}{z_{\mathrm{R}}}\right) - \arctan\left(\frac{l}{z_{\mathrm{R}}}\right) \notag \\
    &\approx \arctan\left(\frac{L+l}{z_{\mathrm{R}}}\right),
\end{align}
where $z_{\mathrm{R}}$ represents the Rayleigh length expressed as
\begin{align}
z_{\mathrm{R}} &= \sqrt{(L+l)(R-L-l)}.
\end{align}
Due to the apex mirror with radius of curvature, the Gouy phase shift from MC3 to MC1 becomes $\eta_{\mathrm{MC3\to MC1}} = 2\times \eta_{\mathrm{MC3\to MC2}}$.
By using the relationships between the beam size at waist and Rayleigh length expressed as
\begin{align}
    w_0 &= \sqrt{\frac{z_{\mathrm{R}}\lambda}{\pi}},
\end{align}
and the trigonometric functions and inverse trigonometric functions described as
\begin{align}
    \sin(\arctan(x)) &= \frac{x}{\sqrt{1+x^2}},
\end{align}
we can calculate the real part of each normalized coefficient as
\begin{align}
    \mathcal{R}\{C_1\} &= R\left(g+\frac{2l}{R}\right)\theta_3, \\
    \mathcal{R}\{C_2\} &= Rg\theta_3, \\
    \mathcal{R}\{C_3\} &= R\theta_3.
\end{align}
One can compute other beam spot response as the same manner.
Hence, the beam shift on each mirror surface due to MC3 pitch misalignment can be described as Eqs. (\ref{pit_beam}) and (\ref{yaw_beam}).

\section{Observable DoFs of aLIGO IMC}
We will consider the observable DoFs of the IMC when the beam is injected to the longer side of the IMC such as aLIGO \cite{Mueller2016}.
In this case, the WFS signals in pitch can be expressed as
\begin{align}
\begin{pmatrix}
	\mathrm{tilt} \\
	\mathrm{trans.}\\
\end{pmatrix}_{\mathrm{Pitch}}
&= \sqrt{2}\gamma w_0
\begin{pmatrix}
	g\gamma & 0 & g\gamma \\
	-g & -\sqrt{2} & -g \\
\end{pmatrix}
\begin{pmatrix}
	\psi_1 \\
	\psi_2 \\
	\psi_3
\end{pmatrix}.
\end{align}
This sensing matrix can be decomposed by SVD as the same manner as short-side-injection case.
The decomposed sensing matrix of pitch can be written as
\begin{align}
\hat{S}_{\mathrm{Pitch}} &= 
I
\begin{pmatrix}
\sigma_1 & 0 & 0 \\
0 & \sigma_2 & 0 \\
\end{pmatrix}
\begin{pmatrix}
	1 & a & -1 \\
	-g & b & -g \\
	1 & 0 & -1
\end{pmatrix},
\label{aLIGO_pit}
\end{align}
where $\sigma_{1,2}$ are singular values of sensing matrix.
Here we used $a$ and $b$ since these elements are 
In the case of aLIGO IMC, the SVD form sensing matrix is more complicated than the case of KAGRA-like configuration and the observable DoFs are difficult to interpret.
On the other hand, one can discriminate the DoF we cannot sense corresponds to the beam waist tilt.

In order to know the observable DoFs, the unitary rotation matrix is adopted, $R(\theta)$, which satisfy following relation
\begin{align}
    R(\theta)_{\mathrm{Pitch}}
    \begin{pmatrix}
	1 & a & -1 \\
	-g & b & -g \\
	1 & 0 & -1
    \end{pmatrix}
    &=
    \begin{pmatrix}
	\mathrm{trans.} \\
	\mathrm{MC2} \\
    \mathrm{tilt}
    \end{pmatrix}.
\end{align}
By introducing this rotation matrix to Eq. (\ref{aLIGO_pit}), the decomposed sensing matrix can be expressed as
\begin{align}
\hat{S}_{\mathrm{Pitch}} &= 
I
\begin{pmatrix}
\sigma_1 & 0 & 0 \\
0 & \sigma_2 & 0 \\
\end{pmatrix}
R^{T}(\theta)_{\mathrm{Pitch}}
\begin{pmatrix}
	\mathrm{trans.} \\
	\mathrm{MC2} \\
    \mathrm{tilt}
\end{pmatrix}.
\end{align}
One can see that we can sense the beam waist translation and MC2 spot position by the WFS.

As the same manner as pitch case, for the case of yaw, we can obtain the SVD form sensing matrix with rotation matrix as
\begin{align}
\hat{S}_{\mathrm{Yaw}} &= 
I
\begin{pmatrix}
\sigma_1 & 0 & 0 \\
0 & \sigma_2 & 0 \\
\end{pmatrix}
\begin{pmatrix}
	1 & a_{\mathrm{Yaw}} & -1 \\
	-g & b_{\mathrm{Yaw}} & -g \\
	1 & 0 & -1
\end{pmatrix},
\end{align}
Similarly to the case of pitch, we can define the unitary rotation matrix such that
\begin{align}
    R(\theta)_{\mathrm{Yaw}}
    \begin{pmatrix}
	1 & a' & -1 \\
	-g & b' & -g \\
	1 & 0 & -1
    \end{pmatrix}
    &=
    \begin{pmatrix}
	\mathrm{trans.} \\
	\mathrm{MC2} \\
    \mathrm{tilt}
    \end{pmatrix}.
\end{align}
Therefore, the sensing matrix for yaw can be decomposed as
\begin{align}
\hat{S}_{\mathrm{long, Yaw}} &= 
I
\begin{pmatrix}
\sigma_1 & 0 & 0 \\
0 & \sigma_2 & 0 \\
\end{pmatrix}
R^{T}(\theta)_{\mathrm{Yaw}}
\begin{pmatrix}
	\mathrm{trans.} \\
	\mathrm{MC2} \\
    \mathrm{tilt}
\end{pmatrix}.
\end{align}

Thus, the observable DoFs are consisted by beam waist translation and MC2 beam spot motion.
Since the IMC has much smaller short side than longer side, the observable composed by beam waist translation and MC2 beam spot can be approximately regarded as the motion of longer side axis of the IMC.
Therefore, one possible interpretation of observable of aLIGO IMC is the shift of longer side axis.% which correspond to the beam axis of the IMC output.

\section*{Funding}
This work was supported by MEXT, JSPS Leading-edge Research Infrastructure Program, JSPS Grant-in-Aid for Specially Promoted Research 26000005, JSPS Grant-in-Aid for Scientific Research on Innovative Areas 2905: JP17H06358, JP17H06361 and JP17H06364, JSPS Core-to-Core Program A. Advanced Research Networks, JSPS Grant-in-Aid for Scientific Research (S) 17H06133, the joint research program of the Institute for Cosmic Ray Research, University of Tokyo, National Research Foundation (NRF) and Computing Infrastructure Project of KISTI-GSDC in Korea, Academia Sinica (AS), AS Grid Center (ASGC) and the Ministry of Science and Technology (MoST) in Taiwan under grants including AS-CDA-105-M06, the LIGO project, and the Virgo project.

\section*{Acknowledgements}
This paper has JGW Document No. JGW-P2011199.

\section*{Disclosure}
The authors declare no conflicts of interest.

% Bibliography
\bibliography{asc}

% Full bibliography added automatically for Optics Letters submissions; the following line will simply be ignored if submitting to other journals.
% Note that this extra page will not count against page length
\bibliographyfullrefs{asc}

\end{document}